\documentstyle[preprint,eqsecnum,aps]{revtex}

\begin{document}
\preprint{DFPD 95/TH/19}
\draft

\title{Markov diffusions in comoving coordinates \\
and stochastic quantization \\
of the free relativistic spinless particle}

\author{Laura M. Morato}
\address{Facolt\`a di Scienze dell' Universit\`a di Verona,
Via delle Grazie, 37134 Verona, \\
and Dipartimento di Fisica "G. Galilei",
Universit\`a di Padova, \\
Via Marzolo 8, 35131 Padova, Italy}
\author{Lorenza Viola}
\address{Dipartimento di Fisica "G. Galilei", Universit\`a di Padova,
and INFN, \\
Sezione di Padova, Via Marzolo 8, 35131 Padova, Italy}

\date{\today}

\maketitle

\begin{abstract}
 We revisit the classical approach of comoving coordinates in
relativistic hydrodynamics and we give a constructive proof for their
global existence under suitable conditions which is proper for
stochastic quantization. We show that it is possible to assign stochastic
kinematics for the free relativistic spinless particle as a Markov
diffusion globally defined on ${\sf M}^4$. Then introducing dynamics by
means of a stochastic variational principle with Einstein's action, we are
lead to positive-energy solutions of Klein-Gordon equation. The
procedure exhibits relativistic covariance properties.
\end{abstract}

\pacs{02.50.-r, 03.65.-w, 03.30.+p}

\section{INTRODUCTION}

Probability Theory, and Markov processes in particular,
are currently recognized as key elements for a proper modelization of
various relativistic phenomena within the framework of the Euclidean
formulation of Quantum Field Theory. On the other hand, the construction
of a mathematically consistent description of diffusive phenomena
in Minkowskian spaces must face severe problems. This is due essentially
to the difficulty to make the Markovian property compatible with the
required relativistic covariance (for different approaches and proposals
concerning this longstanding open problem see for example Refs.
\cite{dudley,hakim,ruggiero,dohrn,serva,vigier,angelis,angelis1} and
\cite{nelson1}). A satisfactory theory seems to be still lacking.

The aim of this work is to introduce a new method for treating phenomena
described by Markovian diffusions in the configurational space
within a relativistic setting. As a first
application we procede to perform a covariant stochastic quantization of
the free relativistic spinless particle.
{}From a physical point of view, the idea underlying our construction can
be sketched as follows. Let us consider, for the sake of simplicity,
a three-dimensional Markov diffusion with constant
coefficient equal to $\nu$, satisfying It\^{o}'s stochastic differential
equation
\begin{equation}
dq(t)=b(q(t),t)\,dt+\nu^{\,\frac{1}{2}}\,dw(t)\;,\hspace{1cm}
t \in [0, +\infty)\;,
\label{diff}
\end{equation}
\noindent where $b$ is a drift-field and $w$ a standard Wiener process.
Introducing the time dependent density $\rho$ of the process $q(t)$, one
can show that under some regularity assumption \cite{nelson,carlen} there
exists a velocity field $v$ (usually called the "current velocity field")
such that the following continuity equation holds:
\begin{equation}
\partial_t \rho + \nabla \cdot (\rho\, v)=0\;.  \label{continuity}
\end{equation}
\noindent Thus, if $b$ is sufficiently smooth, to every Markov process
given by a solution of (\ref{diff}) can be associated an "hydrodynamical
(Eulerian) structure" represented by the time dependent couple $(\rho, v)$,
which satisfy the constraint (\ref{continuity}).

Let us suppose that, together with the diffusion, we are interested to
consider, at a generic point $q$ and time $t$, basic observable quantities as
energy-momentum tensor, entropy, temperature and so on. Thus, in
agreement with the approach usually adopted in the relativistic
hydrodynamics of a perfect fluid, we would define those objects in
the frame of an observer which moves at point $q$ and time $t$ with
velocity $v(q,t)$ (the "comoving observer" at point $q$ and time $t$).
Intuitively, it is therefore reasonable to extend this prescription also
to the "probabilistic" objects we are considering. This idea was in fact
already exploited in Ref. \cite{localklein} to
show that it is possible to redefine Nelson's stochastic kinematics in
a "small" neighborhood of properly defined "comoving observers";
starting from this and a suitable extension of the Einstein action,
the Klein-Gordon equation for the free relativistic particle was there
derived without {\it ad hoc} assumptions.

In the present work we will go beyond the local approach outlined above,
and we introduce instead a "global comoving coordinate system".
A non linear transformation is performed on flat space from natural
coordinates to a class of comoving ones. Our discussion will not
depend on the particular choice in this class and the results will be
expressed in the language of general curvilinear coordinates. The method
is intimately connected with classical results as Frobenius
theorem for vector fields and it is close to the construction of cosmic
standard coordinates in General Relativity (see for example Ref.
\cite{generalrel}).
We will give a precise and simple formulation of the basic mathematical
facts which does not seem to be commonly available in the literature and
which, in particular, is proper for the application to stochastic
quantization.

The content of the work is organized as follows: in Sec. II we give a
sufficient condition for a vector field $V$ on ${\sf R}^m,\;m>1$,
induces a "global comoving coordinate system". In Sec. III the same result
is reformulated for vector fields on ${\sf M}^4$ and pseudo-Riemannian
manifolds.
In Sec. IV we apply these results to redefine Nelson's stochastic kinematics
for a free relativistic spinless particle and to derive in a simple and
fully consistent mathematical way the Klein-Gordon equation. The
stochastic interpretation of different subsets of solutions to this
equation is given in Sec. V, where the non relativistic limit is also
described. Finally, in Sec. VI we discuss some possible physical implications
and briefly consider some future developments and applications.

\section{VECTOR FIELDS GENERATING GLOBAL COMOVING COORDINATES \protect\\
IN EUCLIDEAN SPACES}

Let us consider a vector field $V\, : \,{\sf R}^{1+n} \longrightarrow
{\sf R}^{1+n}\;,\; x \mapsto V(x)$. We shall use for the components of
$x$ and $V$ the notation $x\,=\,(x^0,x^1,\ldots,x^n)$ and
$V\,=\,(V^0,V^1,\ldots,V^n)$. We are searching conditions
on $V$ so that there
exists a coordinate transformation $\Phi$ which changes the algebric
vector $V(x)$ into $V'(x):=(\|V(x)\|,0,\ldots,0)$ for every $x \in
{\sf R}^{1+n}$. The inverse function theorem allows to prove the
following:

{\bf Theorem 1:} Let $V$ be a vector field on ${\sf R}^{1+n}$ with
components $(V^0,V^1,\ldots,V^n)$, which satisfies the following
conditions:\\
\noindent
{\it i}$\,$)\ \ \  $V^0(x) \neq 0 \hspace{1cm}\forall x
\in {\sf R}^{1+n}\;,$ \\
\noindent
{\it ii}$\,$)\ \  There exists a scalar field $S \in {\cal C}^r
({\sf R}^{1+n},{\sf R})\;, r \geq 2$, such that
\[
V^{\mu}(x)=\frac{\partial}{\partial {x^{\mu}}}\,S(x)\;, \hspace{1cm}
\forall \mu=0,1,\ldots,n\;,\; \forall x \in {\sf R}^{1+n}\;. \]
\noindent
Then there exists a diffeomorphism $\Phi\,:\, {\sf R}^{1+n}
\longrightarrow {\sf R}^{1+n}$ such that
$${V'}^{\mu}(x)=
\sum_{\nu=0}^n \:\frac{\partial {\Phi}^{\mu}(x) }{\partial x^{\nu}}\,
V^{\nu}(x) \:= \,\left \{ \begin{array}{cl}
                       \|V(x)\| & \hspace{1cm}\mbox{if } \mu=0\;,  \\
                          0     & \hspace{1cm}\mbox{otherwise}\;.
                      \end{array} \right. \nonumber $$

{\it Proof:} Since $V$ is ${\cal C}^1$, then there exists the congruence
$C_V$ generated by $V$. That is, given any point $x\in {\sf R}^{1+n}$,
there exists a unique integral curve $\gamma_x \in C_V$ to which $x$
belongs. For a fixed $x_0 \in {\sf R}^{1+n}$ let us consider the
equation
\begin{equation}
S(x)-S(x_0)\,=\,0\;. \label{implicit}
\end{equation}
\noindent Denoting thus $y_0$ the $n$-dimensional vector
$(x_0^1,\ldots,x_0^n)$, by assumptions {\it i}$\,$) and {\it ii}$\,$) and the
implicit function theorem there exists a neighborhood $I_{y_0} \subset
{\sf R}^n$ of $y_0$ and a function
$f \in {\cal C}^1(I_{y_0},{\sf R})$ such that
\[ S(\,f(x^1,\ldots,x^n),x^1,\ldots,x^n\,)\, - \,S(x_0)=0 \hspace{1cm}
\forall \,(x^1,\ldots,x^n) \in I_{y_0}\;. \]
\noindent
Let now $\Gamma$ denote the border of $I_{y_0}$ and ${\sf p}$ any point
belonging to $\Gamma$. Then the $\lim_{\,{\sf q}\rightarrow {\sf p}}
f({\sf q})$ may be finite or infinite (the existence is ensured by the
fact that the application of the implicit function theorem in ${\sf p}$
would otherwise lead to a contradiction). In the first case one can
apply again the implicit function theorem at ${\sf p}$, and so on. Thus
for each $x_0 \in {\sf R}^{1+n}$ equation (\ref{implicit}) defines a
unique regular $n$-dimensional hypersurface $\Sigma_{x_0}$ in
${\sf R}^{1+n}$. We also observe that the vector
\[ V(x)\,=\,\biggl( \frac{\partial S(x)}{\partial x^0},
\frac{\partial S(x)}{\partial x^1},\ldots,\frac{\partial S(x)}{\partial
x^n} \biggr) \]
\noindent
is by construction orthogonal to the surface in the point $x$. Hence we can
conclude that for each point $x_0$ in ${\sf R}^{1+n}$ there exists a
unique hypersurface $\Sigma_{x_0}$ and a unique integral curve
$\gamma_{x_0}$ to which $x_0$ belongs. In addition, $\gamma_{x_0}$ is
orthogonal to $\Sigma_{x_0}$ in $x_0$. In other words, the congruence
$C_V$ is globally hypersurface-orthogonal.

To construct the new coordinate system in ${\sf R}^{1+n}$ let us fix a
point $O'$ as new origin and consider $\gamma_{O'}$ (with the
natural orientation) and $\Sigma_{O'}$. Being $\Sigma_{O'}$ a regular
hypersurface imbedded in ${\sf R}^{1+n}$, it is also a $n$-dimensional
Riemannian manifold. Let us introduce on it a coordinate system
$q^1,\ldots,q^n$ (notice that this surface has a unique chart ${\sf
R}^n$) and let $\sigma_{ij}\:,\:i,j=1,\ldots,n$ be its metric tensor,
induced by the Euclidean scalar product in ${\sf R}^{1+n}$. Then to
every point $x \in {\sf R}^{1+n}$ there corresponds a unique point on
$\gamma_{O'}$ (intersection between $\gamma_{O'}$ and $\Sigma_x$) with
coordinate $\lambda(x)$ and a unique point on $\Sigma_{O'}$
(intersection of $\Sigma_{O'}$ and $\gamma_x$) with coordinates
$q^1(x),\ldots,q^n(x)$, and viceversa. Therefore the map (see Fig. 1)
\begin{eqnarray}
\Phi \;: \hspace{5mm} {\sf R}^{1+n} & \longrightarrow &{\sf R}^{1+n}\;,
\nonumber \\
    x\hspace{2mm} & \longmapsto & \hspace{2mm} \xi:=
(\lambda(x),q^1(x),\ldots,q^n(x)) \label{comoving}
\end{eqnarray}
\noindent is one-to-one. Existence and continuity of first-order partial
derivatives of $\Phi$ immediately follows from the regularity of the
integral curves and of the hypersurfaces we are considering.

\vspace{0.8cm}

Let us now denote ${\varphi}_x$ the solution of the ordinary differential
equation
\[  \dot{\xi}(\tau) \,=\,V (\xi(\tau)) \;,  \]
\noindent with initial condition
\[  \varphi_x(\tau)\,=\,x\;.   \]
\noindent Hence we have
\begin{eqnarray}
d\xi^{\mu}(\tau) & = & \frac{\partial {\Phi}^{\mu}}{\partial x^{\nu}}\,
d{\varphi}_x^{\nu}(\tau) + O( {\|d\varphi_x(\tau)\|}^2)  \nonumber \\
\mbox{}          & = & \frac{\partial {\Phi}^{\mu}}{\partial x^{\nu}}\,
V^{\nu}(x)\,d\tau + O( {d\tau}^2)\;.   \label{chainrule}
\end{eqnarray}
\noindent Let us now choose $\lambda(x)$ equal to the length of the arc on
$\gamma_{O'}$ and $\xi^0 := \lambda(x)$. Then we observe that defining
${V'}^{\mu}$ by the equality
\begin{equation}
{V'}^{\mu}\,:=\,\frac{\partial {\Phi}^{\mu}}{\partial x^{\nu}}\,
V^{\nu}\;,   \label{vector}
\end{equation}
\noindent we immediately see that, since $d\xi^{\mu}(\tau)=0$ for
$\mu=1,\ldots,n$ and $d\xi^0(\tau)=\|V(x)\|\,d\tau + O(d\tau^2)$, only
the component ${V'}^0$ is different from zero and it is equal to
$\|V(x)\|$.\hfill$\Box$

\vspace{0.8cm}

It is not necessary to particularize the coordinate $\lambda$ to be the
length of the arc on $\gamma_{O'}$. In fact we can prove the following:

{\it Corollary 1:} Let $\xi^0 = \lambda$ be any coordinate on
$\gamma_{O'}$. Introducing on $\gamma_{O'}$ the positive metric
function $f(\lambda)$ we have
$${V'}^{\mu}(x)\, = \,\left \{ \begin{array}{cl}
 \frac{\|V(x)\|}{\sqrt{f(\lambda(x))}} & \hspace{1cm}\mbox{if } \mu=0\;,  \\
                                 0     & \hspace{1cm}\mbox{otherwise}\;.
                      \end{array} \right. \nonumber $$

{\it Proof:} Denoting by $l(\lambda)$ the length of the arc
$(O',\lambda)$, we can rephrase eqs. (\ref{chainrule})-(\ref{vector}) by
observing that
\begin{eqnarray*}
d\xi^0\,=\,d\lambda\,=\,\frac{dl(\tau)}{\sqrt{f(\lambda)}}\,=\,
\frac{\|V(x)\|}{\sqrt{f(\lambda)}}\,d\tau +
O(d\tau^2)\,=\,{V'}^0(\xi)\,d\tau+O(d\tau^2)\;,
\end{eqnarray*}
\noindent from which the result follows.\hfill$\Box$

\vspace{0.8cm}

The coordinate transformation $\Phi$ transforms the Euclidean metric
tensor $\delta_{\mu\nu}$ into a new one, which we denote by
$g^E_{\mu\nu}$. This transformed tensor has in turn a particularily simple
structure, as one can immediately recognize by
observing that any vector $dx \in {\sf R}^{1+n}$ applied at the point
$x$ can be decomposed into two orthogonal vectors $dz$ and $dy$, where
$dz$ is tangent to $\gamma_x$ at $x$ and $dy$ is tangent to $\Sigma_x$
at $x$. Thus we can write
\begin{eqnarray}
\|dx\|^2 & = & \|dz\|^2 + \|dy\|^2  \nonumber \\
         & = & f(\lambda)\,{d\lambda(x)}^2 + \sum_{i,j=1}^n\, \sigma_{ij}
               (q^1(x),\ldots,q^n(x))\,dq^i(x) \, dq^j(x) + O(\|dx\|^4)\;,
\nonumber \end{eqnarray}
\noindent
where as before $f(\lambda)$ and $\sigma_{ij}$ are the components of
the metric tensor of $\gamma_{O'}$ and $\Sigma_{O'}$ in the given coordinate
$\lambda(x)$ and $q^1(x),\ldots,q^n(x)$ respectively. Thus we can state
the following:

{\it Corollary 2:} The global coordinate transformation $\Phi:\, x
\mapsto \xi:=\Phi(x)\;,x\in {\sf R}^{1+n}$, transforms the Euclidean metric
tensor $\delta_{\mu\nu}$ into the tensor $g^E_{\mu\nu}$ with components
$$ g^E_{\mu\nu}(\xi) \: := \,\left \{ \begin{array}{cl}
 f(\xi^0) & \hspace{1cm}\mu=\nu=0\;,  \\
 0  & \hspace{1cm}\mu=0\,,\,\nu=1,\ldots,n\:;\:\nu=0\,,\,\mu=1,\ldots,n\;, \\
\sigma_{\mu\nu}(\xi^1,\ldots,\xi^n) & \hspace{1cm}\mbox{otherwise}\;.
                    \end{array} \right. \nonumber $$

\vspace{0.8cm}

Owing to the explicit structure of the transformed metric tensor, we are
able to completely characterize our construction from the geometric
point of view. In particular, being the map $\Phi$ a global coordinate
transformation on flat space ${\sf R}^{1+n}$, we know that the Riemann
curvature tensor associated to $g^E$ identically vanishes. We shall see
now that this property extends to $\sigma$, so that $\Sigma_{O'}$ eventually
has only an extrinsic curvature. We summarize the result in the following:

{\it Corollary 3:} The hypersurface $\Sigma_{O'}$ has zero intrinsic
curvature.

{\it Proof:} We need to prove that the $n$-dimensional Riemann curvature
tensor $R^{(n)\,i}_{\,klm}\,,$ $i,k,l,m=1,\ldots,n$ associated to $\sigma$
is equal to zero. Starting from the $(1+n)$-dimensional tensor associated
to $g^E$, let us  express it, as usual, in terms of Christoffel symbols:
\begin{equation}
R^{(1+n)\,i}_{\hspace{2mm}\;\:klm} = \frac{\partial \Gamma^i_{km}}
                                          {\partial \xi^l} -
                       \frac{\partial \Gamma^i_{kl}}{\partial \xi^m} +
\Gamma^i_{\alpha l}\Gamma^{\alpha}_{km} -
\Gamma^i_{\alpha m}\Gamma^{\alpha}_{kl}
\hspace{1.5cm}\alpha=0,\ldots,1+n \;.
\label{riemann1}
\end{equation}
\noindent From Corollary 1 and the well known general relation
\[ \Gamma^{\:\alpha}_{\;\mu\nu}\,=\,\frac{1}{2}\,g^{\alpha\beta}
\biggl(\frac{\partial g_{\beta\mu}}{\partial \xi^{\nu}} +
\frac{\partial g_{\beta\nu}}{\partial \xi^{\mu}}-
\frac{\partial g_{\mu\nu}}{\partial \xi^{\beta}}
\biggr) \;, \]
\noindent one can easily check that the connection $\Gamma$ induced from
$g^E$ has a particularily simple structure, which we write as
\[ \Gamma\,=\,\Gamma^{(1)} \oplus \Gamma^{(n)}\;, \]
\noindent
with $\Gamma^{(1)}\,= \,\Gamma^0_{00}$ and
$\Gamma^{(n)}\,= \,\{\Gamma^{\:i}_{jk}\},\:i,j,k=1,\ldots,n$
corresponding to the metrics on $\gamma_{O'}$ and $\Sigma_{O'}$
respectively. Exploiting this fact, eq. (\ref{riemann1}) becomes
\[  R^{(1+n)\,i}_{\hspace{2mm}\;\:klm}\, = \,R^{(n)\,i}_{\,klm} +
\Gamma^i_{0 l}\Gamma^0_{km}-\Gamma^i_{0 m}\Gamma^0_{kl}\,=\,
R^{(n)\,i}_{\,klm} \;. \]
\noindent The left hand side is zero by flatness, so that necessarily
$R^{(n)\,i}_{\,klm}= 0$.\hfill$\Box$

\vspace{0.8cm}

It is also worth to stress that, as usually done in the theory of
Riemannian manifolds, we can introduce normal coordinates
$\xi_{\ast}^0,\xi_{\ast}^1,\ldots,\xi_{\ast}^n$ in a
neighborhood $I_x$ of each point $x \in {\sf R}^{1+n}$. This is obtained
by choosing normal coordinates both in the one-dimensional
neighborhood of the projection of $x$ on $\gamma_{O'}$ and in the
$n$-dimensional neighborhood of the projection of $x$ on $\Sigma_{O'}$
along $C_V$. (Indeed, due to the flatness of $\Sigma_{O'}$ (Corollary
3), this can be done globally, so that it possible to define a coordinate
system that yields an orthonormal coordinate basis at every point of
${\sf R}^{1+n}$). Hence, in case $O'$ coincides with $O$, the linear part
of the transformation ${\Phi}_{\ast}:y \mapsto \xi_{\ast}:= \Phi_{\ast}(y)$
can be thought as a rotation on ${\sf R}^{1+n}$ which aligns the
$x^0$-axis along the direction of the vector field and which is different
from point to point if the latter is not constant.

\section{GLOBAL COMOVING COORDINATES IN MINKOWSKIAN SPACES AND
PSEUDO-RIEMANNIAN MANIFOLDS}

In this section the Minkowski spacetime ${\sf M}^4$ is understood to
be endowed with the standard Euclidean topology. (By the way, in the
literature the so-called "path topology" has also been considered
\cite{zeeman}, nevertheless it looks quite unsuitable in our setting).
Then it is straigthforward to extend Theorem 1:

{\bf Theorem 1$'$:} Let ${\sf M}^4$ be equipped with the Euclidean topology
and let $\eta_{\mu\nu}:=\mbox{diag}(-1,1,1,1)$ the Minkowskian metric tensor.
If $V$ is a vector field on ${\sf M}^4$ with covariant components
$(V_0,V_1,V_2,V_3)$ satisfying the following conditions:\\
\noindent
{\it i}$\,$)\ \ \  $V_0(x) \neq 0 \hspace{1cm}\forall x
\in {\sf M}^4\;,$ \\
\noindent
{\it ii}$\,$)\ \  There exists a scalar field $S \in {\cal C}^r
({\sf M}^4,{\sf R}),\;r \geq 2$, such that
\[
V_{\mu}(x)=\frac{\partial}{\partial {x^{\mu}}}\,S(x)\;, \hspace{1cm}
\forall \mu=0,\ldots,3\;,\; \forall x \in {\sf M}^4\;, \]
\noindent
then there exists a diffeomorphism $\Phi$ such that
$$ {V'}^{\mu}(x):=
\sum_{\nu=0}^n \:\frac{\partial {\Phi}^{\mu}(x) }{\partial x^{\nu}}\,
V^{\nu}(x) \:= \,\left \{ \begin{array}{cl}
  \sqrt{|\,V_{\mu}(x)V^{\mu}(x)\,|} & \hspace{1cm}\mbox{if } \mu=0\;,  \\
                              0     & \hspace{1cm}\mbox{otherwise}\;.
                      \end{array} \right. \nonumber $$

{\it Proof:} With the given topology, ${\sf M}^4$ is a pseudo-Riemannian
manifold with a unique chart ${\sf R}^4$, cartesian natural coordinates
and metric tensor $\eta_{\mu\nu}$. Thus the proof of Theorem 1 remains the
same as far as it does not involve the metric tensor. This occurs only
where we assert that the vector $V(x)$ is orthogonal to $\Sigma_x$ at
point $x$. But we have seen that the Euclidean product between a vector
with (covariant) components $\partial S(x) / \partial x^{\mu}$ and any
vector $\hat{e}(x) \in {\sf R}^4$ which is tangent to $\Sigma_{O'}$ at
point $x$ is equal to zero. Hence we can write
\[ \frac{\partial}{\partial x^{\mu}} S(x)\,\hat{e}^{\mu}(x)=
V_{\mu}(x)\,\hat{e}^{\mu}(x) = \eta_{\mu\nu}\,V^{\mu}
\,\hat{e}^{\nu}(x)=0\;,   \]
\noindent
so that, by the arbitrariness of the tangent vector $\hat{e}(x)$, the
vector $V(x)\,=\,(V^0,V^1,V^2,V^3)$ is orthogonal to $\Sigma_x$ at
the point $x$ in the Minkowskian metric. \hfill$\Box$

\vspace{0.8cm}

We shall now for definiteness suppose that the congruence $C_V$ is timelike.
Corollaries 1, 2 and 3 can be easily extended in the following way.

{\it Corollary 1$\,'$:} Let $\xi^0 = \lambda$ be any coordinate on
$(O',\lambda)$. Then introducing on $\gamma_{O'}$ the negative metric
function $g_{00}(\lambda)$ we have
$${V'}^{\mu}(x)\, = \,\left \{ \begin{array}{cl}
 \frac{\sqrt{-\,V^{\mu}(x)V_{\mu}(x)}}{\sqrt{-g_{00}(\lambda(x))}} &
    \hspace{1cm}\mbox{if } \mu=0\;,  \\
  0     & \hspace{1cm}\mbox{otherwise}\;.
                      \end{array} \right. \nonumber $$

\vspace{0.8cm}

{\it Corollary 2$\,'$:} The global coordinate transformation $\Phi:\, x
\mapsto \xi:=\Phi(x)\;,x\in {\sf M}^4$ transforms the Minkowskian metric
tensor $\eta_{\mu\nu}$ into the tensor $g_{\mu\nu}$ defined by
$$ g_{\mu\nu}(\xi) \:= \,\left \{ \begin{array}{cl}
 g_{00}(\xi^0) & \hspace{1cm}\mu=\nu=0\;,  \\
 0  & \hspace{1cm}\mu=0\,,\,\nu=1,2,3\:;\:\nu=0\,,\,\mu=1,2,3\;, \\
\sigma_{\mu\nu}(\xi^1,\xi^2,\xi^3) & \hspace{1cm}\mbox{otherwise}\;.
                    \end{array} \right. \nonumber $$

{\it Proof:} This structure immediately comes from that of $g^E$. In fact
we can write
\begin{equation}
g^E_{\mu\nu}(\xi)\,=\, \delta_{\alpha\beta}
\frac{\partial \Phi^{-1\,\alpha}}{\partial \xi^{\mu}}(\xi)\,
\frac{\partial \Phi^{-1\,\beta}}{\partial \xi^{\nu}}(\xi)  \label{emetric}
\end{equation}
\noindent and
\begin{equation}
g_{\mu\nu}(\xi)\,=\, \eta_{\alpha\beta}
\frac{\partial \Phi^{-1\,\alpha}}{\partial \xi^{\mu}}(\xi)\,
\frac{\partial \Phi^{-1\,\beta}}{\partial \xi^{\nu}}(\xi) \;, \label{mmetric}
\end{equation}
\noindent so that a comparison between (\ref{emetric}) and
(\ref{mmetric}) yields $g_{\mu\nu}=g^E_{\mu\nu}$ for $\mu \neq 0$ and
$\nu \neq 0$ while
$g_{00}(\xi^0)=-g^E_{00}(\xi^0)=-f(\xi^0)$.\hfill$\Box$

\vspace{0.5cm}

Corollary 3 still holds (the same proof with $n=3$).

\vspace{0.8cm}

Exactly as before, it is possible to introduce normal coordinates
$\xi_{\ast}$ in a neighborhood $I_x$ of any point $x \in {\sf M}^4$ by
projecting $I_x$ on $\Sigma_{O'}$ along $C_V$. In this case, if $O'$
coincides with $O$, the coordinate transformation $\Phi_{\ast}: y \mapsto
\xi_{\ast}:= \Phi_{\ast}(y)\:,\: y \in I_x$ has the linear part which
reduces, up to a spatial rotation, to a Lorentz boost associated to
$V(x)$, or, explicitely:
\[ {\biggl( \frac{\partial \Phi_{\ast}^{\mu}}{\partial x^{\nu}}
\biggr)}(x)\,=\, \Lambda^{-1\,\mu}_{\;\hspace{2.5mm}\nu}(x)
\hspace{2cm} \forall x \in {\sf M}^4\;.  \]

\vspace{0.8cm}

Theorem $1'$ can also be generalized to vector fields defined on a
pseudo-Riemannian manifold ${\sf M}$. In fact, given a vector field
$\hat{V}$ on ${\sf M}$, let $x\,=\,(x^0,x^1,\ldots,x^n)$
denote the local coordinates of a point ${\sf p} \in {\sf M}$ and
$(V_0(x),V_1(x),\ldots,V_n(x))$ the covariant components of $\hat{V}$ at
the same point. Thus, in case for any point ${\sf p} \in {\sf M}$ one has
$V_0(x) \neq 0$ and $V_{\mu}(x)=\partial S(x) / \partial x^{\mu}$ for
some $S \in {\cal C}^r (U_{\sf p},{\sf R})$, $r \geq 2$ and $U_{\sf p}$
denoting the local chart, we can apply Theorem $1'$ (or the analogous for
a metric having a different signature) to get the desired coordinate
transformation on the generic chart $U_{\sf p}$.

A local version of this global result is commonly quoted in General
Relativity textbooks.

\section{STOCHASTIC QUANTIZATION OF THE FREE RELATIVISTIC \protect \\
SPINLESS PARTICLE}

Nelson's stochastic kinematics for a spinless quantum particle is
obtained by promoting the classical configuration variables to a Markov
diffusion with constant coefficient equal to $\hbar/m$, $\hbar$ being
reduced Planck's constant and $m$ the mass of the particle:
\begin{equation}
dq(t)=b(q(t),t)\,dt+{\biggl( \frac{\hbar}{m} \biggr) }^
{\,\frac{1}{2}}\,dw(t)\;,\hspace{1cm} t \in [0, +\infty)\;.
\label{nelsondiff}
\end{equation}

As briefly exposed in Sec. I, the hydrodynamical (or Eulerian) picture
of the diffusion process is given, under appropriate regularity
conditions, in terms of the time dependent density $\rho$ and the current
velocity $v$. The dynamics is then assigned by imposing that the motion
extremizes the mean (regularized) classical action written as a
functional of both fields \cite{nelson}, \cite{guerramorato}.
In the case of a
particle subjected to a scalar potential ${\cal V}$, the time evolution of
$\rho$ and $v$ is described by two coupled partial differential equations:
\begin{eqnarray}
& & \left\{ \begin{array}{l}
\partial_t \,\rho + \nabla\cdot (\rho \,v) = 0 \;, \\
\partial_t \,v + (v\cdot \nabla)\,v + \frac{\hbar^2}{2m^2} \nabla\biggl(
\frac{ \nabla^2 \sqrt{\rho}}{\sqrt{\rho}} \biggr) = - \nabla {\cal V}
\;,     \end{array} \right. \label{dynamics} \end{eqnarray}
\noindent
in which $v$ must be irrotational in all points where $\rho$ is positive.
Introducing therefore a scalar field $S$ such that $v=\nabla S / m$ and
performing the standard substitution
\[  \psi(x,t)\,:=\, \sqrt{\rho(x,t)}\,\exp{\biggl\{ \frac{i}{\hbar}
\,S(x,t)} \biggr\} \;,\]
\noindent
one can transform (\ref{dynamics}) into the Schr\"{o}dinger wave-equation
\begin{equation}
i \hbar\,\frac{\partial \psi}{\partial t} \,=\,\biggl[ \,
-\frac{\hbar^2}{2m}\, \nabla^2 + m\,{\cal V} \, \biggr] \, \psi\;.
\label{schrod}
\end{equation}

Our objective is now to extend such a procedure to the quantization of a
free relativistic spinless particle. This will be achieved by exploiting a
suitable comoving coordinate system and starting, as previously done
\cite{localklein}, from the classical Einstein action. We conjecture that
the quantum evolution equation can be written also in the relativistic
situation in terms of a couple of fields $(\rho,v)$; they are expected to
satisfy a covariant hydrodynamic equation which reduces to (\ref{dynamics})
(for ${\cal V}=0$) in the limit $\|v\|\ll c$. The relevant comoving
coordinate system will be that induced by the four-velocity field $V$
associated to $v$. Notice that the field $V$ enters at this level
as an unknown;
the fact that it fulfills both assumptions of Theorem $1'$ will be verified
{\it a posteriori}. To assign the kinematics we
also need to introduce an evolution parameter independent of the
particular choice of the coordinate $\lambda$ on $\gamma_{O'}$. To this
end let as before $l(\lambda)$ denote the length of the arc
$(O',\lambda)$ of $\gamma_{O'}$ and consider a temporal parameter $t$
defined by $t:=l(\lambda)/ c$. By construction we have
\begin{equation}
dl(\lambda)^2=d\xi^0
d\xi_0=-f(\lambda)\,d\lambda^2=-c^2\,dt^2\hspace{0.5cm},\hspace{0.5cm}
dt=\frac{\sqrt{ -{dl(\lambda)}^2 }}{c} =\frac{1}{c}\,
\sqrt{-g_{00}(\xi^0)}\,d\xi^0\;.   \label{propertime}
\end{equation}
\noindent The physical meaning of this choice is that $t$ coincides with
the proper time of the comoving observer having $\gamma_{O'}$ as its
world-line. In the following we will also identify $O=O'$.

We formulate the stochastic kinematics of a \underbar{free} relativistic
spinless particle as \vspace{0.5cm}follows: \\
\noindent
\begin{tabular}{ll}
$h_1$) & \hspace{0.5cm}\parbox[t]{15cm}{there exists a timelike four-velocity
field $V$ on ${\sf M}^4$, satisfying the assumptions of Theorem $1'$, such
that the evolution of the position of the particle with respect to the
parameter $t$ is described by a Markov diffusion in the comoving coordinate
system associated to $V$; the diffusion coefficient is equal to $\hbar/m$,
$m$ being the rest mass of the particle;} \\
$h_2$) & \hspace{0.5cm}\parbox[b]{15cm}{the current velocity of such
a diffusion is equal to zero.} \\
  &
\end{tabular}

Denoting by $q^i\,,\:i=1,2,3$, the spatial comoving coordinates of the
particle on $\Sigma_0$, the time development of the Markov process is
fully specified by writing
\begin{equation}
dq^i(t)\,=\,\beta_{+}^i (q(t), t)\,dt+
{\biggl(\frac{\hbar}{m}\biggr)}^{\frac{1}{2}}\,G^{\,i}_k(q(t))\,dw^k(t)\;,
\label{dq}   \end{equation}
\noindent
where $w$ is a standard Wiener process on ${\sf R}^3$ and $G^{\,i}_k G^{kj}
=g^{ij}$, $g^{ij}=\sigma^{ij}$ being the components of the (positive-definite)
metric tensor of $\Sigma_O$ (Corollary $2'$). We will also assume
that the functions $\beta^i_{+}$ are sufficiently smooth to allow the
existence of the so-called "backward representation" of the process
\cite{nelson,carlen}, namely
\begin{equation}
dq^i(t)\,=\,\beta_{-}^i (q(t), t)\,dt+
{\biggl(\frac{\hbar}{m}\biggr)}^{\frac{1}{2}}\,G^{\,i}_k(q(t))\,
dw^k_{\ast}(t)\;, \label{dqrev}   \end{equation}
\noindent
where $w_{\ast}$ is a reversed standard Wiener process.

By introducing the infinitesimal forward and "backward" increments
\begin{eqnarray}
& & \begin{array}{l}
d^{+}q^i(t)\,:=\,q^i(t+dt)-q^i(t)   \\
d^{-}q^i(t)\,:=\,q^i(t)-q^i(t-dt) \;,
    \end{array} \label{infincrements} \end{eqnarray}
\noindent
we can construct the two four-vectors
\begin{eqnarray}
& & \begin{array}{l}
d^{+}\xi^{\mu}\,:=\, (\,d\lambda,\, d^{+}q^i(t)\,)    \\
d^{-}\xi^{\mu}\,:=\, (\,d\lambda,\, d^{-}q^i(t)\,)
    \end{array} \label{increments} \end{eqnarray}
\noindent
in terms of which we define the \underbar{mean regularized invariant}
\begin{eqnarray}
E\{d^{+}\xi^{\mu}\,d^{-}\xi_{\mu}\} & = &
E\{d^{+}q^i(t)\,d^{-}q_i(t)\}-c^2 dt^2  \nonumber \\
  & = & E\{ v^i_{+}(q(t))\,v_{-\,i}(q(t))\}\,dt^2 - c^2dt^2\;.
\label{invariant}
\end{eqnarray}
\noindent Here eq. (\ref{propertime}) has been used and $v_{+}$, $v_{-}$
are to so-called "invariant drifts" introduced in Refs. \cite{nelson} and
\cite{aldrovandi}; they are vector fields on $\Sigma_O$ defined by
\begin{eqnarray*}
& & \begin{array}{l}
v^i_{+}\,:=\, \beta^i_{+} + \frac{\hbar}{2m}\,g^{jk}\,\Gamma^i_{jk} \\
v^i_{-}\,:=\, \beta^i_{-} - \frac{\hbar}{2m}\,g^{jk}\,\Gamma^i_{jk}\;,
    \end{array} \end{eqnarray*}
\noindent
$\Gamma^i_{jk}$ being the Christoffel symbols associated to $g_{ij}$
(metric connection).

As usual in Stochastic Mechanics, we introduce the current velocity $\beta$
and the osmotic velocity $u$ by combining the drifts:
\begin{eqnarray}
& & \begin{array}{l}
\beta \,:=\, \frac{1}{2}\bigl( v_{+}+v_{-} \bigr) \,=\, \frac{1}{2}
\bigl( \beta_{+} + \beta_{-} \bigr)  \\
u     \,:=\, \frac{1}{2} \bigl( v_{+}-v_{-} \bigr) \;.
    \end{array} \label{velocities} \end{eqnarray}
\noindent It is convenient to introduce a smooth invariant measure on
the manifold $\Sigma_O$ and define the relative covariant probability density
of the process, $\rho(q,t)$, through the equality
\begin{eqnarray*}
E\{F(q(t))\} := \int_{{\sf R}^3} F(q)\,\rho(q,t) \,\sqrt{|g|}\,d^3q\;,
\hspace{1cm} |g|:=\,\mbox{det}[\,g_{ij}]\;,
\end{eqnarray*}
\noindent where $F(q)$ is any Lebesgue integrable function of the process. It
is well known \cite{nelson} that, in case $\rho$ is strictly
positive, $\beta,\,u$ and $\rho$ are connected by the following relations:
\begin{mathletters}
\label{generallabel}
\begin{equation}
u^i\,=\, \frac{\hbar}{2m}\,\nabla^i \ln \rho
\label{osmotic}
\end{equation}
\begin{equation}
\partial_t\, \rho + \nabla_i (\rho \, \beta^i) \,=\,0 \;,
\label{current}
\end{equation}
\end{mathletters}\par\noindent
where $\nabla_i$ denotes the covariant derivative along direction $q_i$
(recall that by construction $\nabla_i \,g_{jk} = 0\,\forall i,j,k$).
For the extension to the case when $\rho$ has zeroes see \cite{carlen}.
According to $h_2$), we should also put $\beta$ equal to zero (so that
$\beta_{\pm}$ become time independent); we prefer to do this later, in order
to show that this choice is naturally enforced by consistency requirements.

Thus, in analogy with the classical case, we consider, for arbitrary events
$a$ and $b$ in ${\sf M}^4$, the following stochastic version of Einstein's
action:
\begin{equation}
A_{[a,b]}\,:=\, -\,mc\, \int_a^b \sqrt{ E\{-\,d^{+}\xi^{\mu}\,d^{-}\xi_{\mu}
\} }\:=\: -\,mc^2 \, \int_{t_a}^{t_b} \sqrt{ 1 - E\biggl\{
\frac{v^i_{+}\,v_{-\,i}}{c^2} \biggr\} } \,dt\;,
\label{action}
\end{equation}
\noindent
which can be written as a functional of $\beta$ and $\rho$ exploiting
(\ref{velocities}) and (\ref{osmotic}). We get
\begin{equation}
A_{[a,b]}\,=\, - \,mc^2\, \int_{t_a}^{t_b} dt\, {\biggl( 1-\frac{1}{c^2}
\int_{{\sf R}^3} (\beta^2-u^2)\,\rho \,\sqrt{|g|}\,d^3q
\biggr)}^{\frac{1}{2}}\;, \label{action1}
\end{equation}
\noindent with the convention $\beta^2 := \beta^i \beta_i =
g_{ij}\beta^i\beta^j$ and similarly for $u^2$. We recall here that,
since the configurational manifold has zero intrinsic curvature, no
Pauli-DeWitt curvature term \cite{dewitt} has to be considered in the
expression of the stochastic action functional.

Following a variational strategy which is analogous to that exploited in non
relativistic Stochastic Mechanics \cite{guerramorato},
we can now extremize the action with respect to independent
variations $\delta \beta$ and $\delta \rho$. Since $\rho$ and $\beta$ are
linked by the continuity equation (\ref{current}), this may be done by
introducing a Lagrangian multiplier $\chi$, so that we construct the modified
action
\begin{equation}
\overline{A}_{[a,b]}\,:=\, A_{[a,b]} + \int_{t_a}^{t_b} dt\, \int_{{\sf
R}^3} \chi\,[\partial_t\,\rho +
\nabla_i(\rho\,\beta^i)]\,\sqrt{|g|}\,d^3q\;.  \label{modified}
\end{equation}
\noindent We require that
\begin{eqnarray}
& & \begin{array}{l}
\delta_v\,\overline{A}_{[a,b]}\,=\,o\,(\delta \beta) \\
\delta_{\rho}\,\overline{A}_{[a,b]}\,=\,o\,(\delta \rho) \\
    \end{array} \nonumber \end{eqnarray}
\noindent
for independent variations $\delta \beta$ and $\delta \rho$ having
compact support on $\Sigma_0 \times [t_a,t_b]$. A straigthforward
calculation yields the two variational equations
\[
\frac{m\, \beta_k}{ \sqrt{ 1- E\biggl\{ \frac{\beta^2}{c^2}-\frac{u^2}{c^2}
\biggr\} } } \,=\, \nabla_k\,\chi \;,\hspace{1cm}k=1,2,3\;, \]
\[
\frac{1}{2}\frac{m\, \beta^2}{ \sqrt{ 1- E\biggl\{ \frac{\beta^2}{c^2}
-\frac{u^2}{c^2} \biggr\} } } +\frac{m}{ \sqrt{ 1- E\biggl\{ \frac{\beta^2}
{c^2}-\frac{u^2}{c^2}\biggr\} } }\, \biggl\{ \frac{1}{2} u^2
+\frac{\hbar}{2m}\, \nabla_iu^i \biggr\} -\frac{\partial \chi}{\partial t} -
\nabla_i\chi\,\beta^i\,=\,0\;.\]
\noindent
We eliminate $\chi$ by taking the gradient of the second equation; we
are left with the following three-dimensional dynamical equation:
\begin{eqnarray}
\frac{\partial}{\partial t} \frac{\beta_k}{ \sqrt{ 1- E\biggl\{ \frac{\beta^2}
{c^2}-\frac{u^2}{c^2} \biggr\} } } & + & (\beta_i\nabla^i) \frac{\beta_k}
{ \sqrt{ 1- E\biggl\{ \frac{\beta^2}{c^2}-\frac{u^2}{c^2} \biggr\} } }
\label{dinameq} \\
& - & \left[ \frac{\hbar}{2m} \frac{ \nabla_i\nabla^i \,u_k}{ \sqrt{
1- E\biggl\{ \frac{\beta^2}{c^2}-\frac{u^2}{c^2} \biggr\} } } +
(u_i \nabla^i) \frac{u_k}{ \sqrt{ 1- E\biggl\{ \frac{\beta^2}{c^2}-
\frac{u^2}{c^2} \biggr\} } } \right]=0\;,
\nonumber  \end{eqnarray}
\noindent
being $\nabla_i\nabla^i$ the ordinary three-dimensional Laplace-Beltrami
operator.

In order to get a covariant expression of (\ref{dinameq}), we need a fourth
equation which in the classical case is obtained by imposing that the
action be stationary with respect to variations of the evolution parameter
and which represents the conservation of energy. In this way the Hamiltonian
is constructed as the conjugate momentum with respect to time. Let us then
consider $t$ as a new dynamical variable $t=t(\tau)$, function of some
invariant auxiliary parameter $\tau$. Since the Lagrangian does not depend
explicitly on $t$, the requirement of stationarity of (\ref{action}) with
respect to smooth variations of $t$ having compact support in
$[\tau_a,\tau_b]$, that is
\[  \delta_t\, A_{[t(\tau_a),t(\tau_b)]}\,=\,o\,(\delta t)\;, \]
\noindent immediately yields
\begin{equation}
\frac{d}{dt}\,\frac{m c^2}{ \sqrt{ 1- E\biggl\{ \frac{\beta^2}{c^2}-
\frac{u^2}{c^2} \biggr\} } } \,=\,0\;.   \label{energy1}
\end{equation}
\noindent
Alternatively we can also identify the Hamiltonian with a first
integral of the motion. If $(\rho_{\ast},\beta_{\ast})$
denotes a given solution of (\ref{dinameq}) and (\ref{current}), then in the
total variation of $\overline{A}$ only the boundary term survive:
\begin{equation}
\delta_{\rho_{\ast},\beta_{\ast}}\overline{A}_{[a,b]}=
\int_{t_a}^{t_b} dt\,\frac{\partial}{\partial t}\int_{{\sf R}^3} \chi
\delta \rho_{\ast} \sqrt{|g|}\,d^3q\;. \label{energya}
\end{equation}
\noindent  We particularize now the variations in the following way:
$$ \begin{array}{l}
\delta \rho_{\ast} = \dot{\rho}_{\ast} \delta t + o(\delta t)  \\
\delta \beta_{\ast} = \dot{\beta}_{\ast} \delta t + o(\delta t)\;,
\end{array}  $$
\noindent
so that expression (\ref{energya}) becomes
\begin{equation}
\delta_{\rho_{\ast},\beta_{\ast}}\overline{A}_{[a,b]}=
\delta t \cdot \int_{t_a}^{t_b} dt\,\frac{\partial}{\partial t}
\int_{{\sf R}^3} (\chi
\dot{\rho}_{\ast})\, \sqrt{|g|}\,d^3q \, +o(\delta t) \;. \label{energyb}
\end{equation}
\noindent  On the other hand, being the dynamics time independent one
can also write
\begin{equation}
\delta_{\rho_{\ast},\beta_{\ast}}\overline{A}_{[a,b]}=
\delta t \cdot \int_{t_a}^{t_b} dt\,
{\biggr(\frac{d}{dt}\overline{L}\biggr)}_{\rho_{\ast},\beta_{\ast}}
+o(\delta t)\;, \label{energyc}
\end{equation}
\noindent
so that, comparing (\ref{energyb}) and (\ref{energyc}) and taking the
limit for $\delta t$ going to zero, we find
\begin{equation}
\frac{d}{dt} \left[ \, \int_{{\sf R}^3} (\chi\,\dot{\rho})\, \sqrt{|g|}\,d^3q
\,-\, mc^2\, \sqrt{1 - E\biggl\{ \frac{\beta^2}{c^2}-\frac{u^2}{c^2}
\biggr\} }  \: \right] \,=\,0\;.    \label{energyd}
\end{equation}
\noindent for any solution $(\rho,\beta)$ of (\ref{dinameq}) and
(\ref{current}) (we note that $\overline{L}= L$ in this case). From this,
continuity equation and an integration by parts give
\begin{equation}
\frac{d}{dt}\, \frac{ mc^2 + E\{mu^2\} }{ \sqrt{ 1-E\biggl\{ \frac{\beta^2}
{c^2}-\frac{u^2}{c^2} \biggr\} } } \,=\,0\;.
\label{energy2}  \end{equation}

As a consequence, we can verify that {\sl a natural assumption in order
the energy be properly defined is that the current velocity
$\beta$ in the comoving coordinates be equal to zero.}

Under this assumption, the density of the process is made time independent
so that $dE\{u^2\}/ dt=0$ by (\ref{osmotic}) and (\ref{energy2}) becomes
identical to (\ref{energy1}). This discussion shows that hypothesis $h_2$)
on the kinematics is in fact \underbar{crucial} in formulating a
mathematically and physically consistent theory.

Putting explicitely hereafter $\beta=0$ and defining the constant
functional
\begin{equation}
\tilde{\gamma}\,:=\, {\biggl( 1+ E\biggl\{ \frac{u^2}{c^2} \biggr\}
\biggr) }^{-\frac{1}{2}}\,= \,\tilde{\gamma}[\rho] \;,
\label{gammatilde}
\end{equation}
\noindent we get the following equations of motion in the comoving
coordinates:
\begin{eqnarray}
& & \left\{ \begin{array}{l}
\frac{\hbar\,\tilde{\gamma}}{2m}\, \nabla_i\nabla^i u_k +
\big( u_i\nabla^i \big)\tilde{\gamma}u_k \,=\,0  \\
\frac{d}{dt}\, \tilde{\gamma}\,mc^2\,=\,0\;.
\end{array} \right. \label{comovingeq} \end{eqnarray}
\noindent
It is apparent that for every fixed solution $u$ of (\ref{comovingeq})
the quantity $\tilde{\gamma}$ acts as a strictly positive constant; in
fact, it drops out from the equations and we can solve the equivalent
problem given by
\begin{eqnarray}
& & \left\{ \begin{array}{l}
\frac{\hbar^2}{2m}\, \nabla_i\nabla^i u_k +
\hbar \big( u_i\nabla^i \big)u_k \,=\,0  \\
\frac{\partial}{\partial \xi^0}\,mc^2\,=\,0\;,
\end{array} \right. \label{comovingeq1} \end{eqnarray}
\noindent
where the original comoving coordinates $(\xi^0,\xi^i)$ are reintroduced
(recall that $\nabla_i =\nabla_{q^i}=\nabla_{\xi^i},\:i=1,2,3$).

In order to express (\ref{comovingeq1}) in a covariant form, one has to
introduce objects with the correct transformation properties. First of all,
let us define a real field $\tilde{p}$ by
\begin{equation}
\tilde{p}(\xi)\,:=\,{\cal N}\,\rho(q)\;, \hspace{1cm}
{\cal N} = {\Biggl( \int_{-\Delta}^{\Delta} \sqrt{-g_{00}(\xi^0)}\,d\xi^0
\Biggr)}^{-1} \;,   \label{ptilde}
\end{equation}
\noindent where the constant ${\cal N}$ has been chosen by demanding
that $\tilde{p}(\xi)$ is a normalized probability density on a fixed
four-dimensional domain  $\Omega :=[-\Delta,\Delta] \times {\sf
R}^3,\,\Delta \in {\sf R}^{+}$:
\begin{equation}
\int_{\Omega} \tilde{p}(\xi)\, \sqrt{|g_{\mu\nu}|}\,d^4 \xi\,=\,
\int_{-\Delta}^{\Delta} {\cal N}\,\sqrt{-g_{00}(\xi^0)}\,d \xi^0
\,\int_{{\sf R}^3} \sqrt{|g|}\,d^3 q\, \rho(q)\,=\,1\;,
\label{ptilde1}
\end{equation}
\noindent with $|g_{\mu\nu|}:=\text{det}[g_{\mu\nu}]$. By construction
$\tilde{p}$ is a density with respect to the invariant measure on $\Omega$,
so that it transforms as a scalar.

As a second step, we introduce also the
two four-dimensional fields with components
\begin{equation}
\tilde{V}^{\mu}\,= \,\left( \frac{c}{ \sqrt{-g_{00}(\xi^0)}}, 0
\right) \hspace{1cm},\hspace{1cm} \tilde{U}^{\mu}\,= \,(0,u^i)\;.
\label{fourvelocities0}
\end{equation}
\noindent Both $\tilde{U}$ and $\tilde{V}$ are four-vectors,
due to different reasons. For the latter the four-vector character
directly follows from the tensorial relation:
\begin{equation}
\tilde{U}^{\mu}\,=\, \frac{\hbar}{2m}\nabla^{\mu} \ln \tilde{p}\;,
\label{osmoticfour}
\end{equation}
\noindent which we obtain from (\ref{osmotic}) and (\ref{ptilde}). The former
is instead the four-velocity field corresponding to the
current velocity vector $\beta$ written in comoving coordinates (where the
spatial part $\beta$ is identically vanishing). It can be noticed that
$\tilde{V}^{\mu}$ satisfy itself a relation similar to (\ref{osmoticfour}):
\begin{equation}
\tilde{V}^{\mu} \,=:\, \frac{1}{m} \nabla^{\mu} \tilde{S}\hspace{1cm},
\hspace {1cm} \tilde{S}(\xi)\,:=\,-mc \int_{-\infty}^{\xi^0} d
\overline{\xi}^0\,\sqrt{-g_{00}(\overline{\xi}^0)}\;.
\label{currentfour}
\end{equation}
\noindent It is important to recognize that eq. (\ref{currentfour})
automatically defines $\tilde{S}$ as a scalar field. In all generality,
the metric function $g_{00}(\xi^0)$ must be negative
definite, normalized according to (\ref{ptilde1}) and sufficiently smooth
to guarantee the existence of the primitive in (\ref{currentfour}).
The simplest choice is obviously to take $g_{00}(\xi^0)= -1$, or,
physically, to adopt the proper time of the comoving observer along
$\gamma_{O}$ as a coordinate, $d\xi^0:= c\,dt$. In this case, the function
$\tilde{S}$ simply reduces to
\begin{equation}
\tilde{S}(\xi)\,=\,-\,mc\,\xi^0\hspace{1cm},\hspace{1cm}\xi^0= c t\;.
\label{essetilde}
\end{equation}

Maintaining the above made positions and exploiting the fact that, by
(\ref{fourvelocities0}), $\tilde{U}^0=0$, $\tilde{V}^i=0,\:i=1,2,3$, we can
now easily rewrite eqs. (\ref{comovingeq1}) as follows
\begin{equation}
g^{\alpha\beta}\,\left[ \,-\, \biggl( \frac{\hbar^2}{2m}
\nabla_{\alpha} \nabla_{\beta}\tilde{U}^{\mu} +
\hbar \tilde{U}_{\alpha} \nabla_{\beta} \tilde{U}^{\mu}
\biggr) \,+ \hbar \tilde{V}_{\alpha} \nabla_{\beta}
\tilde{V}^{\mu} \,\right ]\,=\,0 \;.       \label{covarianteq}
\end{equation}
\noindent The continuity equation can also be rewritten, in terms
of a four-dimensional divergence:
\begin{equation}
\nabla_{\mu}\,[\, \tilde{p} \tilde{V}^{\mu}\,]\,=\,0\;.
\label{covariantcont}
\end{equation}
\noindent Now, by means of the same change of variables exploited to get
the Schr\"{o}dinger equation (\ref{schrod}) from its hydrodynamical
counterpart (\ref{dynamics}), it can be easily realized that
(\ref{covarianteq}) and (\ref{covariantcont}) correspond respectively to
the real and imaginary part of a stationary four-dimensional
Schr\"{o}dinger equation, namely
\begin{equation}
-\frac{\hbar^2}{2m}\,g^{\alpha\beta}
\nabla_{\alpha}\nabla_{\beta} \, \tilde{\phi}\,=\, \mu\, \tilde{\phi}\;,
\label{fourschrod} \end{equation}
\noindent where
\begin{equation}
\tilde{\phi}(\xi)\,:=\, \sqrt{\tilde{p}(\xi)} \,\exp{\biggl
\{ \frac{i}{\hbar} \,\tilde{S}(\xi)}
\biggr\} \;. \label{fitilde}
\end{equation}

We remark at this point that, had we maintained $\tilde{\gamma}$ (eq.
(\ref{gammatilde})) in the derivation, then we would have found an apparently
non linear equation, which is quoted in \cite{localklein}. We observe also
that, as usual in
Stochastic Mechanics, the eigenvalue $\mu$ is equal to the mean
value of an energy. In the particular case one can show by direct
computation that
\begin{equation}
\mu =  \int_{{\sf R}^3 \times [-\Delta,\Delta]} \biggl\{
\frac{1}{2} m \tilde{V}_{\mu}\tilde{V}^{\mu} + \frac{1}{2}m
\tilde{U}_{\mu}\tilde{U}^{\mu}\biggr\}\, \tilde{p}(\xi)\,
\sqrt{|g_{\mu\nu}|}\,d^4\xi \,= \, -\frac{1}{2}\, mc^2
+\frac{1}{2}\, m E\{u^2\}\;.   \label{eigenvalue}
\end{equation}
\noindent
We notice that, since $u$ depends on the quantum state $\tilde{\phi}$
through $\tilde{p}$, the Schr\"{o}dinger-like equation (\ref{fourschrod})
exhibits a spectrum of energy levels. On the other hand, one can also
observe that the ratio of the two contributions in (\ref{eigenvalue}) is
\begin{equation}
\frac{E\{u^2\}}{c^2}\,=\, \frac{\hbar^2}{4\,m^2 c^2}\,
\int_{{\sf R}^3} \nabla_{\mu} \ln \rho \, \nabla^{\mu} \ln \rho \,
\sqrt{|g|}\,d^3q  \;, \label{ratio}
\end{equation}
\noindent so that, avoiding extremely sharp densities, which would
be patological in the free case, one can put with good approximation
\begin{equation}
E\{u^2\}/c^2 \ll 1\;.  \label{threshold}
\end{equation}
\noindent This approximation acquires a clear physical significance if the
right hand side of (\ref{ratio}) is interpreted as the (squared) ratio between
the Compton wavelength $\hbar/mc$ and the typical length scale associated to
the density $\rho$. Converting to energy units, one finds that the typical
energy involved in the description has to be much less than $2\,mc^2$. This
is precisely the energy threshold under which a relativistic single particle
theory is expected to hold. Turning to the four-dimensional Schr\"{o}dinger
equation (\ref{fourschrod}), validity of (\ref{threshold}) means that energy
levels become degenerate and consequently the spectrum collapses to
\begin{eqnarray*}
\mu \approx -\frac{1}{2}\,m c^2 \;.
\end{eqnarray*}
\noindent Incidentally, this relation, also called "Feynman ansatz",
represents a widely accepted working assumption in the literature
\cite{ruggiero}, \cite{feynman}. Equation (\ref{fourschrod}) reduces then to
the Klein-Gordon equation in comoving coordinates:
\begin{equation}
\left[ g^{\alpha \beta}\nabla_{\alpha}\nabla_{\beta}\, -\,
\frac{m^2 c^2}{\hbar^2} \right]\,
\tilde{\phi}(\xi)\,=\,0\;, \hspace{1cm} \xi \in \Omega \;.
\label{kleingord1}
\end{equation}
\noindent
Transforming to cartesian coordinates $\{x^{\mu}\}$ on ${\sf M}^4$, we find
from this the familiar expression
\begin{equation}
\left[ \eta^{\mu \nu}\frac{\partial}{\partial x^{\mu}}
\frac{\partial}{\partial x^{\nu}} \, -\,
\frac{m^2 c^2}{\hbar^2} \right]\,
\phi(x)\,=\,0\;, \hspace{1cm} x \in {\Phi}^{-1}\,(\Omega) \;,
\label{kleingord2}
\end{equation}
\noindent where, from (\ref{fitilde}),
\[  \phi(x)\,:=\, \sqrt{p(x)} \,\exp{ \biggl\{
\frac{i}{\hbar} \,S(x) \biggr\} } =
\sqrt{\tilde{p}(\xi(x))} \,\exp{ \biggl \{
\frac{i}{\hbar} \,\tilde{S}((\xi(x)) \biggr \}} \;. \]
\noindent Since the parameter $\Delta$ which fixes the domain $\Omega$ is
arbitrary, the region where the equation is defined can be sufficiently
large to contain any arbitrarily chosen rectangle in ${\sf M}^4$.

\section{STOCHASTIC INTERPRETATION OF DIFFERENT SUBSETS OF \protect \\
SOLUTIONS TO KLEIN-GORDON EQUATION AND \protect \\ NON RELATIVISTIC LIMIT}

The quantization procedure which we have just outlined naturally selects a
particular subset of the solutions to Klein-Gordon equation. In fact we
observe that, in comoving coordinates,  the conserved four-current density
at an arbitrary point $x \in {\sf M}^4$, defined as
\begin{equation}
\tilde{J}^{\mu}(\xi):=m\,\tilde{p}(\xi)\,\tilde{V}^{\mu}(\xi)=
\tilde{p}(\xi)\nabla^{\mu} \tilde{S}(\xi)\;,\hspace{1cm}\xi=\Phi(x)\;,
\label{comovcurrent}
\end{equation}
\noindent has the following particular structure:
\begin{equation}
\tilde{J}^0(\xi)\,=\,\frac{mc\,\tilde{p}(\xi)}{\sqrt{-g_{00}(\xi^0)}}
\hspace{1cm},\hspace{1cm} \tilde{J}^i(\xi)\,=\,0\;.
\label{structure}  \end{equation}
\noindent We also notice, that the conservation law associated
to (\ref{structure}), namely
\[  \nabla_{\mu}\,\tilde{J}^{\mu}\,=\,0   \]
\noindent is nothing but a rewriting of the continuity equation
(\ref{covariantcont}) for the probability density of the stationary
diffusion $q(t)$.

For a generic solution of the Klein-Gordon equation in the inertial
frame let us correspondly introduce the conserved four-current density by
\begin{equation}
J_{\mu}(x)\,:=\, {| \phi(x) |}^2\, \frac{\partial}{\partial x^{\mu}} S(x)
\,=\,\frac{\hbar}{2\,i}\,\Bigl( \phi^{\ast}\partial_{\mu}\phi-\phi
\, \partial_{\mu}\phi^{\ast} \Bigr) \;.   \label{conservedcurrent}
\end{equation}
\noindent
Denoting by $V(x)$ the four-vector with covariant components $V_{\mu}(x)=
\partial_{\mu} S(x)/m $, we have by construction
\begin{equation}
\tilde{V}^{\mu}(\xi)\,=\, \frac{\partial \Phi^{\mu}}{\partial x^{\nu}}(x)\,
V^{\nu}(x)\;, \hspace{1cm}\xi=\Phi(x)\;.
\label{vtilde}
\end{equation}
\noindent
We then choose in a neighborhood of $\xi$ normal coordinates $\xi_{\ast}:=
(\xi^0_{\ast},\xi^1_{\ast},\xi^2_{\ast},\xi^3_{\ast})$. As observed in Sec.
III, we get
\begin{equation}
\tilde{V}^{\mu}_{\ast}(\xi)\,=\,
\Lambda^{-1\,\mu}_{\hspace{2.5mm}\nu}(x) V^{\nu}(x)\;  ,  \label{boost}
\end{equation}
\noindent where $\tilde{V}^{\mu}_{\ast}$ denote the normal components of
$\tilde{V}$ and $\Lambda^{\mu}_{\nu}$ is, up to a spatial rotation, the
Lorentz boost associated to the three-dimensional velocity
\begin{equation}
v^i\,:=\,c\,\frac{V^i}{V^0}\;,\hspace{1cm}i=1,2,3\;.
\label{3velocity}
\end{equation}

Thus we can conclude that the comoving coordinate transformation $\Phi$,
represented in normal coordinates in the neighborhood of
a point $x$ by $\Phi = \Phi_{\ast}$, acts on four-vectors as a
Lorentz boost associated to the three-velocity $v(x)$ (which is uniquely
determined by the four-velocity $V(x)$).

We also observe that since $J^{\mu}$ is itself a four-vector, then
\begin{equation}
J^{\mu}(x)\,=\,\Lambda^{\mu}_0(\xi)\,\tilde{J}^0_{\ast}(\xi)
\;, \label{jmu}
\end{equation}
\noindent with (from (\ref{structure}))
\begin{equation}
\tilde{J}^0_{\ast}(\xi)\,=\,mc\,\tilde{p}(\xi)\;, \label{j2}
\end{equation}
\noindent Recalling the transformation properties of $\tilde{p}$ and
expliciting the Lorentz boost, one finds from (\ref{jmu})-(\ref{j2})
\begin{equation}
J^0\,=\, \frac{mc\,p}{ \sqrt{1-\frac{v^2}{c^2}} }\hspace{1cm},\hspace{1cm}
J^i\,=\, \frac{mv^i\,p}{ \sqrt{1-\frac{v^2}{c^2}} }\;, \label{j3}
\end{equation}
\noindent so that
\begin{eqnarray}
& & \left\{ \begin{array}{l}
J^0\, \geq\,0   \\
J_{\mu}J^{\mu}\,=\,-\,m^2c^2\,{|\phi|}^2\;.
\end{array} \right. \hspace{1cm}\forall x \in \Phi^{-1}(\Omega)\;.
\label{conditions}
\end{eqnarray}

Thus we can immediately recognize that {\sl a necessary and sufficient
condition for a solution to Klein-Gordon equation corresponds to a
one-particle diffusion satisfying the variational principle in the
comoving coordinates is that conditions (\ref{conditions}) hold}.

In fact, necessity was just proven. Conversely, given a normalized
solution of the Klein-Gordon equation with associated four-current density
$J$, sufficiency immediately comes by putting, in all points where
$J^0 \neq 0$,
\begin{equation}
v^i\,:=\,c\,\frac{J^i}{J^0} \hspace{1cm},\hspace{1cm}
p\,:=\, \phi^{\ast}\phi \;.   \label{sufficiency}
\end{equation}
\noindent We have from this
\begin{eqnarray*}
J_{\mu}J^{\mu}\,=\,-{(J^0)}^2\,\biggl(1-\frac{v^2}{c^2}\biggr)\,=\,
-\,m^2c^2 \,p^2
\end{eqnarray*}
\noindent
and consequently
\begin{equation}
J^0\,=\, \frac{mc\,p}{\sqrt{1-\frac{v^2}{c^2}}}\,=\, \Lambda^0_{\mu}
\tilde{J}_{\ast}^{\mu} \hspace{1cm},\hspace{1cm}
J^i\,=\, \frac{mv^i\,p}{\sqrt{1-\frac{v^2}{c^2}}}\,=\, \Lambda^i_{\mu}
\tilde{J}_{\ast}^{\mu} \;, \label{j4}
\end{equation}
\noindent which proves the assertion.

\vspace{0.8cm}

Some remarks are now in order. First of all, we notice that the set of
solutions restricted by conditions (\ref{conditions}) is not empty since it
contains at least all {\sl positive frequency} plane-wave solutions
(in a suitable limit of non normalized $\phi$). Note also that the
positivity constraint on $J^0$ consistently allows to interpret it as a
conserved density in hydrodynamical sense.

Furthermore, this stochastic quantization procedure can be extended in
an obvious way to the description of a beam of identical (spinless) non
interacting particles: in this case, one would simply recover positive
energy solutions fulfilling both (\ref{conditions}) but such that
\[  \int_{{\sf M}^4} { |\phi| }^2\,d^4x\,=\,N\;,  \]
\noindent $N$ denoting the total number of particles. (This is trivially
obtained by substituting the non normalized density $\rho N$ in place of
$\rho$ at the beginning of our derivation).

A less obvious fact is that it is possible to give a fully probabilistic
interpretation also to solutions with \underbar{negative} energy. The
idea can be sketched as follows. Let us start from a ${\sf R}^3$-valued
diffusion having zero current velocity and satisfying the stationary
stochastic differential equation
\begin{equation}
dq(t)=\beta_{+}(q(t))\,dt +\nu^{\,\frac{1}{2}}\,dw(t)=
\,u(q(t))\,dt+\nu^{\,\frac{1}{2}}\,dw(t)\;,\hspace{1cm}
t \in [0, T]\;,\:dt>0\;,
\label{stocdiff}
\end{equation}
\noindent with $u$ denoting, as usual, the osmotic velocity and $T>0$
arbitrary. Let us now think of a diffusion with a "specular" time evolution.
In order to have an example of a (classical) phenomenon that would be described
by means of such a "specular" diffusion, one could imagine a cloud of brownian
particles which, starting from an initial spreaded spatial distribution, would
concentrate after some interval of time in a small region. This is a "rare"
event to which usual probabilistic models would give zero probability, but
which is not, of course, in principle impossible. To get a mathematical
description in terms of "forward" differentials, let us firstly introduce
time-reversal in the usual way:
\begin{eqnarray}
& & \left\{ \begin{array}{l}
t'=\,-\,t \\
q'(t')=\,q(t) \\
q'(t'-\delta)=\,q(t+\delta)\;,
\end{array} \right.    \label{timereversal}
\end{eqnarray}
\noindent where $\delta$ is a fixed positive time increment. We have
immediately from (\ref{timereversal})
\begin{equation}
dq'(t'-\delta) \,:= \,q'(t')-q'(t'-\delta)\,=\,-\,dq(t) \,=,
-\,( q(t+\delta)-q(t) )\;. \label{reversal1}
\end{equation}
\noindent
But $dq(t)$ is assigned from eq. (\ref{stocdiff}); in particular, we can
use the corresponding so-called "backward representation" to write
\begin{equation}
dq'(t'-\delta) \,=\,-\,dq(t)= \,-\, \{ -\,u(q(t))\,dt+\nu^{\,\frac{1}{2}}
\,dw_{\ast}(t)  \} \;,
\label{backward}
\end{equation}
\noindent where we have exploited the relation $\beta_{-}=- \,\beta_{+}$
holding in this particular stationary case\cite{nelson}, and where
$w_{\ast}$ is a standard reversed Wiener process, that is a process
which has all properties of an usual Wiener process except from the fact
that its increments are independent of the $\sigma$-algebra generated by
the future of $w_{\ast}(t)$. Let us introduce
\begin{equation}
w'(t')-w'(t'-\delta)\,:=\,w_{\ast}(t)-w_{\ast}(t+\delta) \;.
\label{wprimo}
\end{equation}
\noindent From the definitions and the properties of $w_{\ast}$, one can
immediately see that $w'(t')$ is a standard Wiener process (which,
loosely speaking, "goes forward in time"). By inserting eq. (\ref{wprimo})
in eq. (\ref{backward}), we are able to write, after standard manipulations, a
stochastic differential equation for the diffusion $q'$, namely
\begin{equation}
dq'(-t)= u(q'(-t))\,dt+\nu^{\,\frac{1}{2}}\,dw'(-t)\;,\hspace{1cm}
-t \in [-T,0]\;,\;dt>0\;.
\label{reversed}
\end{equation}
\noindent Eq. (\ref{reversed}) provides a possible forward description
of what we have called a "specular diffusion". Now, all of these
considerations can be extended to the stationary kinematics in comoving
coordinates. We can thus conclude that to every solution of the Klein-Gordon
equation with positive $\tilde{J}^0$ there corresponds a solution where $t$ is
replaced by $-t$. The first one is stochastically represented by a
stationary diffusion in the comoving coordinates (with respect to the
invariant parameter $t$), while the second one corresponds to the "rare
event" of a "specular diffusion". Changing the parameter $t$ into $-t$
simply leads to a change of sign in $\tilde{S}$, as one can easily see
recalling eq. (\ref{essetilde}) and the fact that $\tilde{S}$ is a
scalar by construction. Then $\tilde{J}^0$ becomes negative and
consequently
\[ J^0\,=\, \Lambda^0_{\mu} \tilde{J}_{\ast}^{\mu}\,=\,-\,
\frac{ mc\,p}{ \sqrt{1-\frac{v^2}{c^2}} } \]
\noindent is also negative at every point. This furnishes us with the
desired interpretation.

\vspace{0.8cm}

Having clarified how different subsets of solutions to Klein-Gordon
equation can receive different stochastic interpretation, we are now in
position to study the non relativistic limit described by Schr\"{o}dinger
equation. Since the latter represents the evolution equation for the quantum
state of a single standard spinless particle, we need to consider only
those solutions which satisfy conditions (\ref{conditions}); in addition,
we must also select those for which $\|v\| \ll c$. We expect that these last
solutions approximately solve dynamical equations equivalent to
the Schr\"{o}dinger one.

The simplest starting point is the Klein-Gordon equation written in cartesian
coordinates $\{x^{\mu}\}$:
\begin{equation}
\left[ \eta^{\mu \nu}\frac{\partial}{\partial x^{\mu}}
\frac{\partial}{\partial x^{\nu}} \, -\,
\frac{m^2 c^2}{\hbar^2} \right]\,
\phi(x)\,=\,0\;, \hspace{1cm}
\phi(x)\,:=\, \sqrt{p(x)} \,\exp{ \biggl\{
\frac{i}{\hbar} \,S(x) \biggr\} } \;, \label{start}
\end{equation}
\noindent being $p(x)$ a positive normalized probability density over
${\sf M}^4$. If we define
\begin{equation}
V_{\mu}\,:=\, \frac{1}{m}\, \frac{\partial}{\partial x^{\mu}}S(x)\;,
\label{vumu}
\end{equation}
\noindent
then, in agreement also with (\ref{sufficiency}), there exists a velocity
field $v$ on ${\sf R}^3$ such that
\begin{equation}
V^0\,=\, \frac{c}{ \sqrt{1-\frac{v^2}{c^2}} }\hspace{1cm},\hspace{1cm}
V^i\,=\, \frac{v^i}{ \sqrt{1-\frac{v^2}{c^2}} }\;. \label{v2}
\end{equation}
\noindent
We also construct the four-dimensional osmotic velocity by
\begin{equation}
U_{\mu}\,:=\, \frac{\hbar}{2m}\, \frac{\partial}{\partial x^{\mu}}\ln
p(x) \;.   \label{u1}
\end{equation}
\noindent The Klein-Gordon equation (\ref{start}) can be now expressed
in terms of $U$ and $V$: by following, in reversed order, the same steps
leading from eqs. (\ref{covarianteq}) and (\ref{covariantcont}) to eq.
(\ref{fourschrod}), the real part and the gradient of the imaginary part
constitute a pair of four-dimensional hydrodynamical equations:
\begin{eqnarray}
& & \left\{ \begin{array}{l}
-\,\Bigl( \frac{\hbar^2}{2m}\,\eta^{\mu\nu} \frac{\partial}
{\partial x^{\mu}} \frac{\partial}{\partial x^{\nu}} U +
\hbar U^{\mu}\frac{\partial}{\partial x^{\mu}} U \Bigr) +
\hbar V^{\mu}\frac{\partial}{\partial x^{\mu}}
V\,=\,0   \\
\frac{\partial}{\partial x^{\mu} }\,(p\,V^{\mu}) \,=\,0\;.
\end{array} \right. \label{hydrodin}
\end{eqnarray}

Starting from (\ref{hydrodin}), we now claim that {\sl the non relativistic
limit is recovered, after substituting $V^0$ and $V^i$ in terms of
(\ref{v2}), by taking the limit $\|v\|/c \ll 1$ in the resulting expression,
having care to treat as negligible all terms which are multiplied by
$\hbar^2/c$ and $\hbar^2/c^2$.}

The proof goes as follows: let us denote by $\hat{u}$ the spatial part of
$U$ and by $v$ the three-dimensional velocity field associated
to $V$, and let $\hat{t}$ be the time parameter in the inertial frame
associated with $\{x^{\mu}\}$. We put, as usual, $dx^0:= c\,d\hat{t}$. Letting
$\|v\|/c \ll 1$, the four-dimensional continuity equation can immediately be
rewritten as
\[ \partial_t\, p + \nabla\cdot( p\,v)\,=\,0\;.\]
\noindent
The second hydrodynamical equation is now considered in two steps.
Separation of the spatial part leads to the following three-dimensional
equation:
\begin{eqnarray}
-\,\biggl[ \frac{\hbar}{2m} \nabla^2 \hat{u}-\frac{\hbar}{2m}
\frac{\partial^2 \hat{u}}{\partial {x^0}^2} & + & (\hat{u} \cdot  \nabla)
\hat{u} +
\Bigl( U^0 \frac{\partial}{\partial x^0} \Bigr)\hat{u}
\biggr]  \nonumber \\
  & + & \biggl( \frac{v}{\sqrt{1-\frac{v^2}{c^2}}} \cdot \nabla \biggr)
\frac{v}{\sqrt{1-\frac{v^2}{c^2}}} + \biggl( \frac{c}
{\sqrt{1-\frac{v^2}{c^2}}} \frac{\partial}{\partial x^0} \biggr)
\frac{v}{\sqrt{1-\frac{v^2}{c^2}}}\,=\,0 \;.  \label{spatial}
\end{eqnarray}
\noindent
If, as before, we rewrite $dx^0= c\,d\hat{t}$ and take into account the
explicit expression of $\hat{U}$ (\ref{u1}), we can neglect the second and
fourth term, both of which depend on $\hbar^2/c^2$. In the limit $\|v\|
\ll c$, Eq. (\ref{spatial}) becomes then
\begin{eqnarray*}
-\,\biggl[ \frac{\hbar}{2m} \nabla^2 \hat{u} + ( \hat{u}\cdot \nabla )
\,\hat{u} \biggr] + ( v \cdot \nabla ) \,v +
\frac{\partial v}{\partial \hat{t}} \,=\,0 \;.
\end{eqnarray*}
\noindent This, together with the continuity equation, is nothing but the
hydrodinamical form of Schr\"{o}dinger equation, which we have also quoted
in (\ref{dynamics}). The temporal component of the dynamical equation
(\ref{hydrodin}) gives rise instead to the following one-dimensional relation:
\begin{eqnarray}
-\,\biggl[ \frac{\hbar}{2m} \nabla^2 U_0 -\frac{\hbar}{2m}
\frac{\partial^2 U_0}{\partial {x^0}^2} & + &
\Bigl( U^{0} \frac{\partial}{\partial x^{0}} \Bigr)U_0 +
\Bigl( U^{i} \frac{\partial}{\partial x^{i}} \Bigr)U_0
\biggr]  \nonumber \\
  & + & \biggl( \frac{v}{\sqrt{1-\frac{v^2}{c^2}}} \nabla \biggr)
\frac{c}{\sqrt{1-\frac{v^2}{c^2}}} + \biggl( \frac{c}
{\sqrt{1-\frac{v^2}{c^2}}} \frac{\partial}{\partial x^0} \biggr)
\frac{c}{\sqrt{1-\frac{v^2}{c^2}}}\,=\,0 \;.  \label{temporal}
\end{eqnarray}
\noindent
The last two terms are readily recognized to be neglegible in the limit
$\|v\|/c \ll 1$. Recalling the explicit form of $U_0$ coming from
(\ref{u1}), we see that the other terms depend on $\hbar^2/c$. This proves
the assertion.

\vspace{0.8cm}

It is worth to note that the quantum non relativistic limit appears as an
approximation of the correct relativistic dynamical equations from a
twofold point of view: firstly, the approximation $\|v\| \ll c$ corresponding
to the classical non relativistic limit is done; secondly, terms depending on
$\hbar^2/c$ are also neglected. This fact is perhaps a little
unexpected and conceptually intriguing.

Furthermore, an important point of this approach needs to be stressed here,
namely the fact that {\sl in the relativistic framework no diffusion process
is supposed to exist for the inertial observer}. A simple probabilistic
interpretation is consequently lost in a generic inertial frame. This
difficulty may be considered, in a sense, as the stochastic counterpart
of non trivial questions which, in the framework of canonical
quantization, are related to a consistent definition of observable
quantities, such as position operator (see \cite{angelis} and
\cite{schweber} for any detail). In our procedure, the reason for this
fact is that, as observed above, the Wiener process has sample paths which
can "go outside" the light cone, so that there exist inertial frames where
such spacelike trajectories would appear as going forward and backward
in time.

On the other hand, such a peculiarity disappears in the non
relativistic limit. As a consequence, the usual Nelson's quantization
becomes consistent with the proposed procedure in a non relativistic setting.

\section{DISCUSSION AND OUTLOOK}

The first observation to be done is that stochastic quantization of the
free spinless relativistic particle based on comoving (local or global)
coordinates is very clean both from a mathematical and a physical point
of view. In fact, on one side all probabilistic objects are well
defined; on the other side, once given stochastic kinematics in the
comoving coordinates, dynamical equations naturally emerge from a
stochastic version of the Einstein action, by means of the stationary
action principle. No additional {\it ad hoc} assumptions are necessary,
at variance with the path integral approach \cite{feynman} and previous
attempts within stochastic frameworks which we have already quoted.

We also observe that the stochastic quantization procedure proposed in
this work gives only to the subset of solutions to Klein-Gordon equation
with positive energy and normalized four-current the physical meaning of
representing the quantum evolution of \underbar{one} single free relativistic
spinless particle. As is well known, this is a consistency property which is
not easily achieved within standard canonical (first) quantization
\cite{schweber}.

On the other side it may be interesting that one can give a
probabilistic interpretation also to other solutions of Klein-Gordon equation,
in particular to those with negative energy in terms of specular
diffusions as "rare events".

It may also be worthwhile to insist on the fact that the comoving coordinates
approach allows to more properly understand how to handle covariant and non
covariant quantities. Dynamical equations (which do not contain
stochastic terms) are Lorentz covariant, while
Markovian kinematics is not: in fact, the latter is covariant only with
respect to reparametrizations on $\Sigma_{O}$ and $\gamma_{O}$.

We point out that global comoving coordinates approach introduces a
description of the motion of the particle which is well defined on the
whole ${\sf M}^4$ (this in particular provides the technical advantage
of eliminating all spurious boundary terms which come out in the local
derivation \cite{corrige}). Moreover, it directly leads to a dynamical
theory which is covariant with respect to {\sl arbitrary} changes of
space-time coordinates.

Indeed the results presented in this paper look suitable for the
extension to the description of a quantum particle subjected to a
gravitational (classical) field. In this case one can furthermore
conceive that, in presence of strongly peaked gravitational fields, the
degeneracy among the energy levels of the four-dimensional dynamical
equation (\ref{fourschrod}) breaks down. This is an aspect of the
outlined stochastic quantization procedure which deserves further
investigation \cite{next}.

{}From another point of view, it would be also appealing both on
mathematical and physical grounds to consider a
sort of {\sl second} stochastic quantization, by reinterpreting the
probability density in the comoving coordinates as a physical scalar
field in a space-time dependent random medium. This could in fact be
done by exploiting suitable probabilistic techniques (see Ref.
\cite{molchanov} for an excellent review of this subject).

\acknowledgments

We wish to thank Roberto Onofrio for very useful remarks.


\begin{references}
\bibitem{corrige} We notice that the eigenvalue $\mu$ of eq.
(\ref{fourschrod}) as calculated in the local approach \cite{localklein}
would contain a spurious boundary term which can be made arbitrarily
small by suitably restricting the considered neighborhood.
\end{references}
\end{document}